# HBN-encapsulated, graphene-based room-temperature terahertz receivers with high speed and low noise

Leonardo Viti,[1] David G. Purdie,[2] Antonio Lombardo,[2] Andrea C. Ferrari,[2] Miriam S. Vitiello[1]

1. NEST, Istituto Nanoscienze - CNR and Scuola Normale Superiore, Piazza San Silvestro 12, 56127 Pisa, Italy
2. Cambridge Graphene Centre, University of Cambridge, Cambridge CB3 0FA, UK

**Abstract**

Uncooled Terahertz (THz) photodetectors (PDs) showing fast (ps) response and high sensitivity (noise equivalent power (NEP) < nW/Hz$^{1/2}$) over a broad (0.5 THz – 10 THz) frequency range are needed for applications in high-resolution spectroscopy (relative accuracy ~ 10$^{-11}$), metrology, quantum information, security, imaging, optical communications. However, present THz receivers cannot provide the required balance between sensitivity, speed, operation temperature and frequency range. Here, we demonstrate an uncooled THz PD combining the low (~2000 $k_B\mu m^{-2}$) electronic specific heat of high mobility (> 50000 cm$^2$V$^{-1}$s$^{-1}$) hBN-encapsulated graphene with the asymmetric field-enhancement produced by a bow-tie antenna resonating at 3 THz. This produces a strong photo-thermoelectric conversion, which simultaneously leads to a combination of high sensitivity (NEP≤160 pWHz$^{-1/2}$), fast response time (≤ 3.3 ns) and a four orders of magnitude dynamic range, making our devices the fastest, broadband, low noise, room temperature THz PD to date.

**Introduction**

Room-temperature (RT) detection over the terahertz (THz) frequency range is of great interest for a number of applications in biomedicine [1,2], security [3], spectroscopy [4], environmental monitoring [5], real-time imaging [6] and high data-rate communications [7], as well as for unveiling fundamental properties of condensed matter systems at the nanoscale [8-10].

Many different RT detection technologies have been developed over the past two decades, with a largely variable range of sensitivities, response times ($\tau$), operational frequency range and underlying physical mechanisms [11]. Commercial RT THz sensors include thermal devices, such as pyroelectric [11] and Golay cells [11], semiconductor-oxide [12] or metallic based micro-bolometers [1] and solid-state electronic architectures, such as Schottky diodes [13] and complementary metal–oxide semiconductor (CMOS) based field effect transistors (FET) [14].

The sensitivity of a THz photodetector (PD) can be expressed in terms of its noise equivalent power (NEP), which indicates the minimum incident optical power required to achieve a unitary signal-to-noise ratio over a bandwidth of 1 Hz [15]. Pyroelectric PDs have NEP~100 pWHz$^{-1/2}$ [11], $\tau$ ~ 10 ms [11], broadband operation over the range 0.2 – 30 THz [11], and are mostly single-pixel devices [6]. Microbolometric THz cameras are the most common multipixel sensors [12]. They have broadband operation (0.2 – 100 THz) [11], low NEP ~ 20 pWHz$^{-1/2}$ [11], but $\tau$ limited to ~ 10-1000 μs [11]. Solid-state electronic devices are significantly faster than these. E.g., Schottky diodes have $\tau$ ≥ 5 ps [16] with NEP ~ 100 pWHz$^{-1/2}$ [16], but their performances rapidly decrease with increasing operational frequency > 3 THz [13], or when implemented in an array configuration [17]. CMOS-based FETs are best suited for multi-pixel integration [18], broadband operation up to 9 THz with NEP ~ 10 pWHz$^{-1/2}$, and with $\tau$ < 1 µs were reported [19,20].

A promising route to combine the main advantages provided by the aforementioned technologies relies in the exploitation of layered materials (LMs). For these, the dominant detection mechanism can be tailored by design [21]. Their ultrafast dynamics [22] and the ease of fabrication [23,24] and integration [25], can boost both the sensitivity and speed of THz PDs operating at RT. Single Layer Graphene (SLG) and other LMs have been used to fabricate a variety of THz PDs [26-36]. RT THz PD in an unbiased FET was demonstrated exploiting overdamped plasma waves (PW) [14,37], photothermoelectric (PTE) rectification [33,36], bolometric detection [38] or via a combination of the aforementioned phenomena [15,26,30,34,38].

The low (~ 100 Ω) channel and contact resistances in SLG FETs (GFETs) help in reducing the detector noise [39]. Contact resistance < 100 Ωµm can be obtained with edge contacts to encapsulated SLG in hexagonal boron nitride (hBN) [40] or with contact area cleaning and rapid thermal annealing [41]. Moreover, when THz rectification is mediated by the simultaneous modulation of carrier density ($n$) and drift velocity ($v_d$) in the channel, *i.e.* in the PW driven response, an increase in carrier mobility ($\mu$) leads to a reduction of $\tau$ [42]. This enables modulation frequencies > 10 GHz in the low-field limit (i.e. as long as velocity saturation effects can be neglected [42]), since the maximum modulation frequency is expected to be proportional to $\mu$ [42].

SLG is also ideal for PTE PDs [36,43], owing to its gapless nature that allows broadband absorption from the UV to GHz frequencies [44]. The PTE effect entails a thermal gradient within the electronic distribution [43], which yields the diffusion of carriers away from the hottest region [44]. In SLG, when electrons are heated up by photon absorption, photogenerated carriers remain thermally decoupled from the crystal lattice [43,45]. This is due to the difference between the electron-electron scattering time (~ 20 fs [22,46], needed for thermalization of the electronic distribution) and the slower (~ 2 ps [43,47]) electron-phonon relaxation time. Therefore, a *quasi*-equilibrium state is reached, where the electronic temperature, $T_e$, is considerably higher than the lattice temperature $T_L$ [45]. The electronic subsystem shows a record-low specific heat $c_e$ (~ 2000 $k_B$µm$^{-2}$ at 300 K, where $k_B$ is the Boltzmann constant) [48,49], which can lead to the ultrafast (50 fs) onset of thermal gradients [49,50] and to a rapid overheating of the electronic distribution with respect to the SLG lattice [49]. This is ideal for PTE-based devices, since all the absorbed electromagnetic energy is first transferred to electron heating before being lost through other (slower) thermalization channels: interaction with acoustic phonons occurs on a picosecond timescale [22,50]. Therefore, the PTE conversion can be efficient, even though the Seebeck coefficient is relatively small ($S_b$ ~ 40 µVK$^{-1}$ [51]) with respect to other 2D materials (*e.g.* black phosphorus [33]). This is particularly effective at THz frequencies, because of the inhibition of ultrafast relaxation (< 150 fs) via optical phonon emission by photo-excited electrons at energies lower than ~ 0.2 eV [22,46,52,53].

The small $c_e$ stems from the density of states shrinking in proximity of the charge neutrality point (CNP) as a consequence of the linear band dispersion [49]. In particular, the analytical expression of the electronic specific heat $c_e$ in proximity of the CNP (chemical potential $< k_B T$) reads [49]: $c_e = 18\zeta(3)k_B^3 T_e^2 /\pi(\hbar v_F)^2$, where $\zeta(3) = 1.202$ is the zeta function, $\hbar$ is the Planck constant and $v_F = 1.1 \times 10^6$ ms$^{-1}$ is the Fermi velocity. Therefore, $c_e$ grows quadratically with $T_e$, reaching ~2000 $k_B\mu m^{-2}$ at 300 K [48,49]. In contrast, the lattice specific heat ($c_p$) is > 1000 times larger [48]. Hence, the combination of $c_e/c_p < 1000$ with $\mu > 50000$ cm$^2$V$^{-1}$s$^{-1}$ makes GFETs ideal for fast PTE THz PDs. Since SLG with $\mu > 70000$ cm$^2$V$^{-1}$s$^{-1}$ can be produced over large area by encapsulating chemical vapor deposition (CVD) SLG in hBN [54], large area multi-pixel architectures at THz frequencies are feasible.

Single pixel, RT broadband GFET detectors with NEP ~ 80 pWHz$^{-1/2}$ [36] or $\tau$ ~ 0.1 ns have been already reported [28,29,55]. However, in both Refs. [28,29], the PD ultrafast and broadband response was associated with a quite poor NEP > 1 mWHz$^{-1/2}$ in Ref. [28] and > 8 nWHz$^{-1/2}$ in Ref. [29]. These NEP are larger than those required for a practical exploitation of THz RT PDs, especially for imaging, high-resolution spectroscopy and near-field microscopy where NEP < 1 nWHz$^{-1/2}$ is preferable [6]. This motivates the effort to devise SLG PDs combining fast $\tau$ (~ 1 ns), broadband operation (0.1 – 10 THz), large (> 3 orders of magnitude) dynamic range and low (< 1 nWHz$^{-1/2}$) NEP.

In Ref. [36] a PTE SLG PD was reported with NEP < 100 pWHz$^{-1/2}$, $\tau$ ~ 40 ns, a 3 orders of magnitude dynamic range, and operating over a 0.3-4 THz bandwidth. This employed a dual-gated narrow gap (100 nm) dipolar antenna, which, while creating a *p-n* junction in the SLG channel, concentrates the THz field at the junction, where the photoresponse arises [36]. However, the ~ 40 ns response time still hinders the application in pulse characterization or high repetition-rate detection.

Here, we increase the speed (electronic bandwidth) and dynamic range of RT SLG PDs, exploiting a much simpler architecture than Ref. [36], relying on the on-chip patterning of a broadband bow-tie antenna ($\delta\omega/\omega_0 > 20\%$ [56]) to couple THz radiation to a sub-wavelength hBN-SLG-hBN heterostructure with $\mu$ ~ 53000 cm$^2$V$^{-1}$s$^{-1}$. By exploiting both PW and PTE mechanisms we get a low-noise (NEP ~ 160 pWHz$^{-1/2}$) RT THz PD with a four orders of magnitude dynamic range and $\tau$ ~ 3 ns, *i.e.* one order of magnitude faster than any other low NEP (< 10$^{-9}$ WHz$^{-1/2}$) THz PD operating at RT reported so far to the best of our knowledge. A SLG is encapsulated within hBN, forming a clean hBN-SLG-hBN heterostructure [57]. hBN and SLG flakes are prepared by micromechanical exfoliation [58] on intrinsic Si+285nm SiO$_2$ wafers. The flakes are then picked up sequentially (top hBN, SLG, bottom hBN) with polydimethylsiloxane (PDMS) and polycarbonate (PC) stamps. The stack is then released at 180°C on the final Si/SiO$_2$ wafer. This T is higher than the glass transition T of PC (~ 150°C [59]), enabling a better control during transfer thanks to the decreased viscosity of PC [54,57]. Blisters of trapped contaminants at the interface between hBN and SLG become mobile at this T [54,56] and can be pushed along the heterostructure until they reach an edge, leaving the interfaces contaminant-free [54,56].

Once the heterostructure is placed on the substrate, the FET channel is defined by dry etching, leaving the edge of the SLG channel exposed [40,57]. The channel has a rectangular shape with length L$_C$ = 5.4 μm and width W$_C$ = 0.8 μm (Figure 1a). One-dimensional edge contacts are then realized by electron beam lithography (EBL) followed by metal deposition (Cr/Au, 10/100 nm). The source (*S*) and drain (*D*) electrodes are asymmetrically shaped to favor the required asymmetry for which either PTE or PW, or their combination take place within the SLG channel (Figure 1a) [34]. Before defining the top-gate (*G*) electrode, a thin oxide layer ($Al_2O_3$, 10 nm) is deposited via atomic layer deposition (ALD). This prevents leakage current between *G* and the SLG through the edges of the channel itself. The *G* contact, covering a length $L_G$ = 5 μm over the SLG channel, is then patterned by EBL and finalized by metal deposition (Cr/Au, 10/100 nm). Similar to the *S* electrode, the *G* contact is shaped as the branch of a planar bow-tie antenna, with radius $r_b$ = 21 μm and flare angle 90° (Figure 1a).

The antenna dimensions are chosen following electromagnetic simulations with Comsol Multiphysics (see Supplementary Information S1 for further details): a 3 THz radiation, matching the frequency of a THz quantum cascade laser (QCL), impinges on GFET on $Si/SiO_2$ integrated within a planar bow-tie antenna, whose radius $r_b$ is changed in discrete steps from 6 to 69 μm. Figure 1b plots the simulated enhancement of the *in-plane* electric field component ($E_{//}$) provided by the antenna as a function of $r_b$ with respect to the case where the hBN-SLG-hBN heterostructure is not connected to the antenna. The antenna response shows two maxima at $r_b$ = 20 and 52 μm, corresponding to the $\lambda/2$ and $3\lambda/2$ resonances. The inset of Figure 1b shows the *out-of-plane* component of the electric field on the plane of the antenna. The maximum $|E_{//}|^2$ enhancement is 3500, concentrated in the gap between *S* and *G*, creating the required asymmetry [34] for the activation of PW and PTE effects.

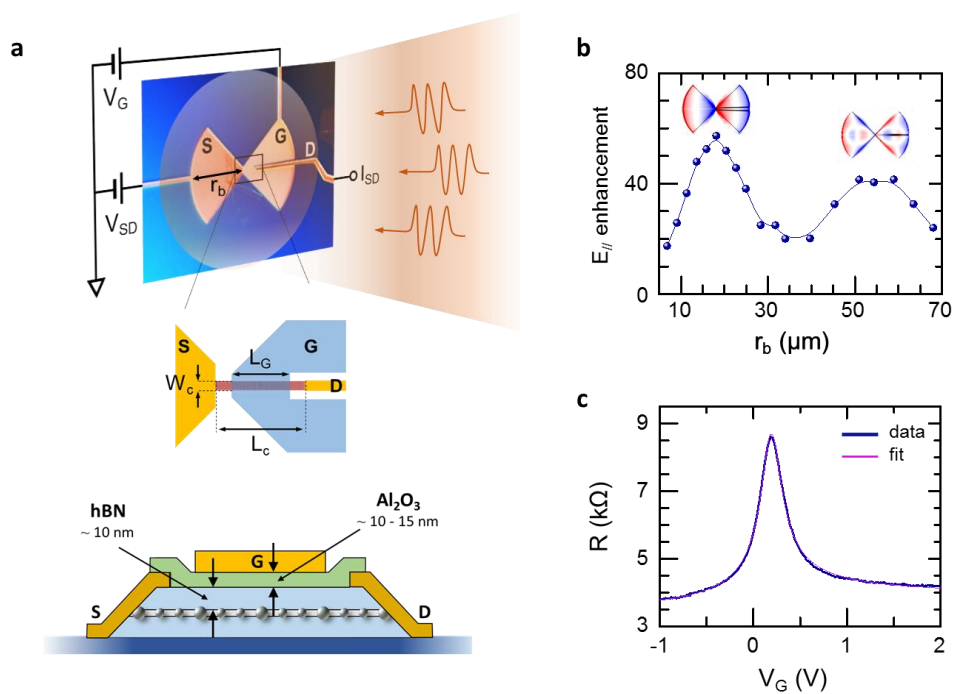

**Figure 1**: PD layout. (a) Bottom: schematic of PD active area. SLG is encapsulated between two flakes of hBN (bottom 30 nm, top 10 nm). The heterostructure is capped by a ~10-15 nm $Al_2O_3$ layer after edge-contact fabrication [40]. Top: detector layout. The GFET is embedded in a planar bow-tie antenna (radius $r_b$ = 21 μm). The inset shows the main geometrical parameters of the GFET: channel width ($W_c$ = 0.8 μm), channel length ($L_c$ = 5.4 μm), gate length ($L_G$ = 5 μm). (b) Antenna simulations showing the enhancement of the *in-plane* component of the electric field $E_{//}$ at the position of the GFET due to the presence of the antenna, plotted as a function of $r_b$, for an impinging frequency of 3 THz. Inset: maps of the out-of-plane component of the electric field for $r_b$ = 20 μm and $r_b$ = 52 μm, showing $\lambda/2$ and $3\lambda/2$ resonances. (c) RT two-terminal resistance as a function of top-gate voltage ($V_G$) from which μ is extracted.

The devices are then electrically characterized at RT. Figure 1c is the channel resistance (R), extrapolated by probing the source-drain current ($I_{SD}$) as a function of top-gate voltage ($V_G$), keeping the source-drain voltage $V_{SD}$ = 2 mV. The CNP is at $V_G$ = 0.2 V. The R($V_G$) plot can be used to extract $\mu_{FE}$, $n_0$ and the contact resistance ($R_0$), by fitting R with [60] R = $R_0$ + ($L_C/W_C$)(1/$n_{2d}e\mu_{FE}$), where $n_{2d}$ is the gate-dependent charge density, given by [60] $n_{2d} = [n_0^2 + (C_{Ga}/e (V_G-CNP))^2]^{1/2}$. Here, $C_{Ga}$ is the gate-to-channel capacitance per unit area ($C_{Ga}$ = 0.2 µFcm$^{-2}$) and CNP is used as fixed parameter for the fitting function. We get $\mu_{FE}$(holes) ~ 41000 ± 800 cm$^2$V$^{-1}$s$^{-1}$, $\mu_{FE}$(electrons) ~ 53000 ± 400 cm$^2$V$^{-1}$s$^{-1}$, $n_0$ ~ 1.52 ± 0.01 x 10$^{11}$ cm$^{-2}$, $R_0$ ~ 3.3 ± 0.01 kΩ and 4.0 ± 0.01 kΩ for hole and electron doping, respectively. These $\mu$ are consistent with Ref. [57], and are the highest reported to date in any THz GFET, to the best of our knowledge.

The GFET is then optically tested using a single-plasmon 2.8 THz QCL, operating at a heat sink T=30 K in a tabletop Stirling cryostat (model Ricor K535). The QCL is driven in pulsed mode (pulse width 1.6 µs; repetition rate 40 kHz). The average QCL output power is progressively varied from a few nW to 820 µW, at the corresponding maximum lattice T=170 K (estimated assuming a substrate thermal resistance ~ 20 K/W [61]). The 30° divergent THz beam is collimated and focused by using two picarin (tsupurica) lenses with focal lengths 25 and 50 mm (Figure 2a). The resulting Gaussian beam at the focal point has a waist ~120 µm. The average optical intensity is increased up to a maximum of ~ 2 W/cm$^2$. The SLG PD is then mounted onto a roto-translation stage, to move it over the focal plane and modify the relative angle ($\alpha$ in Figure 2a) between the bow-tie antenna axis and the vertically polarized THz electric field.

Optical measurements are then performed at RT and at liquid nitrogen T (77 K). The sample is electrically connected as follows: the *S* electrode is grounded, the *G* contact is connected to a *dc* voltage generator (Keithley 2400), and the generated photovoltage singnal *Δu* is measured at the *D* electrode, connected to a lock-in amplifier (Stanford Research 830, reference/modulation frequency $f_{ref}$ = 1.334 kHz) through a voltage pre-amplifier (FEMTO HVA200, gain 100, bandwidth 200 MHz). For the 77K measurements, the sample is mounted on the cold unit of a gas-refrigerated cryostat (QMC, TK 1800) and the THz beam reaches the PD through a 2 mm polymeric window (TPX, transmission 76 ± 2% at 3 THz). *Δu* is then estimated from the photovoltage recorded with the lock-in $V_{LI}$, as *Δu* = 2.2· $V_{LI}/\eta$ [32], where $\eta$ is the voltage preamplifier gain coefficient.

Figure 2b shows the map of |*Δu*| (log scale) at RT on the focal plane (*xy* in Figure 2a) of the THz beam, for an impinging average power ~100 µW. The GFET has a signal-to-noise ratio (SNR) > 10$^3$ at the optimal $V_G$ = 0.36 V.

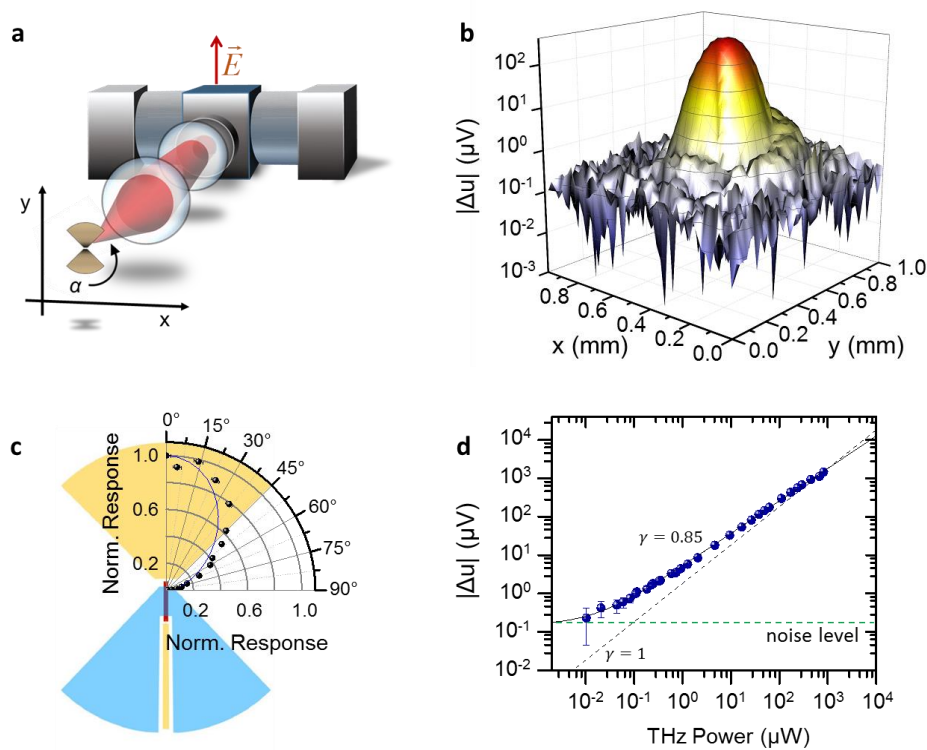

**Figure 2**: Optical characterization. (a) Schematics of the THz experiment: a 2.8 THz QCL is focused on the GFET, whose position (*xy* plane) and orientation with respect to the laser polarization (angle *α*) can be controlled. (b) 1x1 mm $|\Delta u|$ map for an impinging power ~100 μW. The map is obtained by scanning the detector position on the focal *xy* plane and recording the measured photovoltage when $V_G$ = 0.36 V. The ratio between $\Delta u$ measured at the center (*x* = 0.5 mm, *y* = 0.5 mm) and $\Delta u$ measured outside the THz beam is more than three orders of magnitude. (c) Polar plot of the detected signal as a function of *α*, ranging from 0° (antenna axis parallel to the light polarization) to 90° (antenna axis perpendicular to the light polarization). Black dots: experimental data. Solid blue line: simulation. (d) Absolute value of the photovoltage versus incident power in log-log scale ($V_G$ = 0.36 V). The solid line is a fit to the data using $|\Delta u(P)| \sim a_0 + P^\gamma$, where $a_0$ is the experimental noise floor and the exponent $\gamma$ = 0.85. The dashed black line is for $\gamma$ = 1. The error bars are the root means square deviations of the measured $|\Delta u|$.

In order to verify the polarization selectivity of our antenna geometry [56], we measure the THz photoresponse as a function of angle *α* between antenna axis and THz beam polarization. The photoresponse (Fig. 2c) reaches its maximum when the antenna axis is parallel to the polarization and decreases when *α* is increased from 0° to 90°. The experimental data (black dots) are in good agreement with simulations (solid blue line).

An important figure of merit for a THz PD is the dynamic range [32,36], *i.e.* the range of impinging optical power that the PD is capable to sense. To determine it, we vary in regular steps the average output power of the QCL from 0 to 820 μW (Figure 2d, and Supplementary Information figure S2). The GFET detects a minimum power ~ 90 nW and a maximum power ~ 820 μW. The dependence of the response with respect to power is almost linear over more than three orders of magnitude (setup limited), following a power law $|\Delta u| \sim P^\gamma$ with $\gamma$ = 0.85 ± 0.007 (the fit to the data is reported in Figure 2d, black solid line). This quasi-linear dependence of the THz photoresponse is expected for both PW and PTE-based PD operating in the *weak-heating* regime [36,47], *i.e.* when the thermal gradient along the GFET channel is smaller than the heat sink T: $\Delta T \ll$ 300 K [35]. The small deviation from the linear ($\gamma$ = 1) power dependence can be ascribed to the temperature dependence ($\sim T^{-1}$) of graphene thermal conductivity at RT [48].

In order to identify the dominant physical mechanism governing the photodetection process, the PD response is then recorded as a function of $V_G$ at 77 K and 300 K. The responsivity $R_v$ is evaluated by normalizing the photovoltage $\Delta u$ with respect to the optical power impinging on the detector: $R_v = \Delta u/P \cdot A_{spot}/A_{diff}$ [33], where P is the total THz power, $A_{spot}$ is the beam spot area and $A_{diff}$ is the diffraction limited area, calculated as $\lambda^2/4$ [32,33], Fig. 3a (left vertical axis). At RT a maximum $|R_v| = 49$ VW$^{-1}$ is found for $V_G$ = 0.36 V. At 77 K, $|R_v|$ reaches ~180 VW$^{-1}$. Both at 300 and 77 K, the $R_v$ plot as a function of $V_G$ shows a double sign switch. Unlike a purely or a dominant overdamped PW (resistive self-mixing) regime, we do not see a single sign change in $\Delta u$ at the CNP, caused by the sign change in the derivative of the static channel conductance $\sigma$, according to [34]: $\Delta u_{PW} \propto -\sigma^{-1} \partial\sigma/\partial V_G$.

The double sign change in the $R_v$ *vs.* $V_G$ plot can be interpreted as the fingerprint of a dominant THz-light induced PTE [36,43]. In this case, $\Delta u_{PTE} = (S_G - S_u) \cdot \Delta T$ [33, 34], where $S_G$ is the Seebeck coefficient of the SLG below the *G* electrode, $S_u$ is the Seebeck coefficient of the ungated regions close to the *S* and *D* contacts, and $\Delta T$ is the T difference between the *S*-side and the *D*-side of the GFET channel. This thermal imbalance is direct consequence of the asymmetric funneling of THz radiation by the bow-tie antenna. The PTE response is given by the diffusion of hot carriers from the hot (*S*) towards the cold (*D*) side of the GFET, which results in a measurable electrical signal. $S_G$ can be evaluated from the *dc* conductivity $\sigma$ of the GFET, using the Mott equation [33]: $S_{Mott} = -eL_0T \cdot \sigma^{-1}(\partial\sigma/\partial V_G) \cdot (\partial V_G/\partial E_F)$, where $L_0 = (\pi k_B)^2/(3e^2)$ is the Lorenz number and $E_F$ is the Fermi energy. $\partial V_G/\partial E_F$ can be evaluated from $E_F = \hbar v_F(\pi C_{Ga}\delta V_G/e)^{1/2}$ [62], where $\delta V_G = |V_G - V_{CNP}|$. At RT, $S_G$ reaches a maximum of 130 µVK$^{-1}$ for $V_G$ = 0.36 V (see Supplementary Information figure S4). $S_u$ is expected to be ~ $S_G$ when $V_G$ = 0 V [33], therefore it is positive and constant with respect to $V_G$. $S_G - S_u$ is plotted in Fig. 3b for T = 300 K and T = 77 K; in both cases, the two sign changes as a function of $V_G$ are expected. The PTE model can also provide a quantitative interpretation of the increase in $R_v$ at T = 77 K, with respect to RT. Indeed, $\Delta T$ is expected to increase at low T, due to the increased electron cooling length [43,47] and to the improved electrical characteristics of the GFET [35] (details in Supplementary Information). The comparison between the calculated $\Delta u_{PTE}$ at RT and at 77 K is given in Figure 3a (right vertical axis). The maximum PTE response at 77 K is expected to be 3 times larger than the maximum obtained at RT, in quantitative agreement with our measurements.

A more rigorous interpretation can be given by considering the simultaneous interplay of PW and PTE. Figure 3c plots $\Delta u$ at RT as a function of $V_G$ obtained for an optical power of ~100 µW (orange curve), together with the estimated $\Delta u_{PTE}$ and $\Delta u_{PW}$. The solid black line represents the combined theoretical photovoltage $\Delta u_T = \Delta u_{PTE} + \Delta u_{PW} = a \cdot (S_G - S_u) + b\sigma^{-1}(\partial\sigma/\partial V_G)$ where *a* and *b* are the fitting parameters. The PW contribution is only relevant close to CNP, whereas PTE dominates at higher carrier densities ($|V_G - V_{CNP}| > 0.3$ V). From the separate evaluation of the two contributions, and the knowledge of *a* and *b,* we can estimate $\Delta T$ driving the PTE response. By dividing $\Delta u_{PTE}$ (Fig. 3c) by $(S_G - S_u)$ at RT (Fig. 3b), we get $\Delta T = \Delta u_{PTE} / (S_G - S_u)$ ~ 0.8 K when the THz power is ~100 µW, which results in a T gradient ~0.2 Kµm$^{-1}$ along the SLG channel. This confirms that the PD operates in the weak-heating regime [36,47] for all the investigated range of optical THz power (0 – 820 µW).

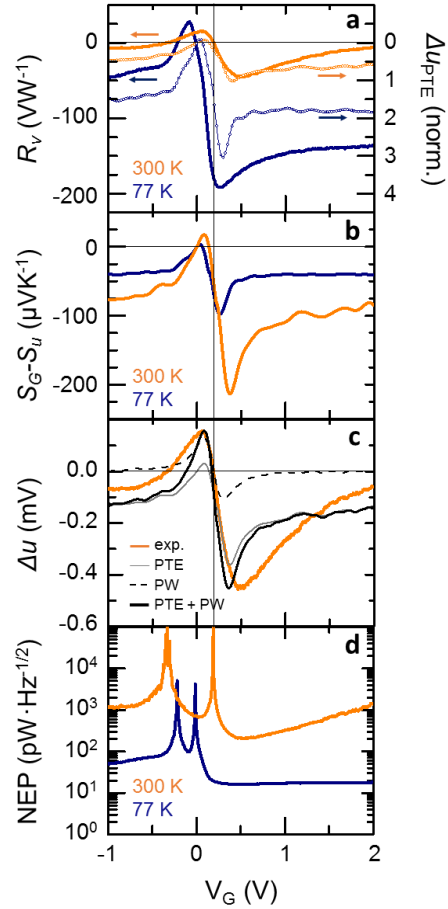

**Figure 3:** Main figures of merit. (a) Left vertical axis, solid lines: $R_V$ vs. $V_G$ at RT and 77 K. The double sign switch in the photovoltage is a signature of a dominant PTE. Right vertical axis, dotted lines: estimated $\Delta u_{PTE}$, normalized to the maximum value calculated at RT. The grey vertical line at $V_G = 0.2$ V indicates the CNP at RT. (b) Estimated difference in the Seebeck coefficient between gated ($S_G$) and ungated area ($S_u$) as a function of $V_G$ at RT and at 77 K. (c) Comparison between experimental $\Delta u$ (measured at P = 100 μW) and theoretical PTE, PW and combined $\Delta u_{PTE} + \Delta u_{PW}$ photovoltages. (d) NEP at 300 K and at 77 K evaluated by assuming a Johnson-noise dominated noise spectral density. Minimum NEPs~ 160 pWHz$^{-1/2}$ and 18 pWHz$^{-1/2}$ are obtained at 300 K and 77 K.

In order to assess the PD sensitivity, we evaluate NEP as the ratio between PD noise spectral density (NSD) and $R_v$. A correct evaluation of NSD is extremely important for a proper estimate of NEP. In our system there are four mechanisms that can have a role in the total noise figure: the Johnson-Nyquist noise ($N_J$) [36], the shot noise [63], the generation-recombination noise [64] and the flicker noise (1/$f$, or telegraph noise) [65]. The first is related to the thermal voltage fluctuations ($V_{th}$) at the ends of the GFET channel and its power spectral density is, in turn, related to R and to the heat sink T via [15] $N_J^2 = <V_{th}^2> = 4k_BTR$. In our case, the $N_J$ contribution to NSD is ~10 nVHz$^{-½}$ at 300 K and ~4 nVHz$^{-½}$ at 77 K.

The shot noise of a quantum conductor typically increases under THz illumination due to the possibility of photon-assisted shot noise (PASN) phenomena [63]. However, in our T range and under zero-bias (no external $V_{SD}$ is applied), the contribution of the shot noise to the total noise figure is expected to be orders of magnitude lower than $N_J$ [63], therefore it can be neglected. The same argument applies to the generation-recombination noise, whose amplitude drops below $N_J$ under zero-bias and for current densities <

This is the authors' version of the submitted article, published in its final form at http://dx.doi.org/10.1021/acs.nanolett.9b05207

1 µA/mm [64]. The 1/$f$ noise can be neglected with respect to $N_J$ due to the combination of zero-bias detection (no direct current applied) and > kHz modulation frequency ($f_{ref}$ = 1.334 kHz) [39]. Thus, we approximate NSD ~ $N_J$. The NEP is then calculated as $N_J/|R_v|$ and the resulting NEP($V_G$) plots are in Fig. 3d for T = 300 and 77 K. We get minimum NEP ~160 and 18 pWHz$^{-1/2}$ for 300 and 77 K, respectively. Notably, the RT NEP is minimum when the GFET is *n*-type, which corresponds to the regime where a *p-n* junction is established between the *S* and *G* electrodes, *i.e.* the region where the antenna funnels THz radiation.

Finally, we evaluate the bandwidth (BW) of our THz PD by taking advantage of the employed QCL. When the QCL is driven with high-voltage pulses ($V_{QCL}$ > 29 V), it enters the so-called negative differential resistance (NDR) regime [66]. From an electrical point of view, this corresponds to a very unstable high field domain regime, in which the driving current fluctuates randomly under increased applied bias. From an optical point of view, the QCL average output power progressively decreases when increasing the voltage, due to the increased overall temperature of the laser lattice, which, in turn, reduces the population inversion [67]. In the NDR, the QCL turns off and on many times during a single pulse, as in Figures 4a-4b. These abrupt transitions are an intrinsic property of the QCL and are governed by the exchange of energy between electrons, photons and lattice within the QCL cavity. Thus, the switching from the *off* to the *on* state (and vice versa) is not dictated by external circuitry (power supply and pulse generator), and can be significantly faster than the onset of externally driven pulses (the switching time is expected to be ~1 ns [67]).

Figures 4a-4b show the time trace of the current flowing through the QCL ($I_{QCL}$) during a single 1.6 µs pulse, recorded with an oscilloscope (resolution 5 GS/s, corresponding to 200 ps) and the corresponding voltage time trace at the output of the GFET, collected at $V_G$ = 0 V, by using a voltage pre-amplifier (Model A1423, CAEN) with input impedance 50 Ω, gain 46 dB and bandwidth 1.2 GHz. The instantaneous $P_i$ switches on and off in an almost periodic way, with a period of 210 ns. In coincidence with the *off*-state condition, $I_{QCL}$ shows pronounced dips (indicated by vertical arrows in Fig. 4a), ascribed to the sudden reduction in the current flow due to the unstable transport arising from the *off-on* turning state within the QCL active region.

The waveform in Figure 4b is used to assess the detector BW. Figure 4c shows a zoom on a single $P_i$ oscillation, from which the rise-time $\tau_{on}$ and fall-time $\tau_{off}$ are extracted using the fitting functions [36] $V_{out} = c_0 + V_{on} \cdot [1-\exp(-(t-c_1)/\tau_{on})]$ and $V_{out} = c_2 + V_{off} \cdot \exp(-(t-c_3)/\tau_{off})$. The fitting parameters $c_0$, $c_2$ and $c_1$, $c_3$ are constants representing voltage offsets and time offsets, respectively; $V_{on}$ and $V_{off}$ are the voltage jumps in the waveform. We obtain a rise (fall) time ~ 3.3 ns (4.2 ns), corresponding to BW ~ 53 MHz (38 MHz), where BW = $1/2\pi\tau$, one order of magnitude better than what was observed in Ref. [36] for a comparable NEP. The GFET itself is expected to show response time of the order of 100 ps [28,29,36]. Then the observed rise and fall times are limited by external factors. The first limitation is represented by the detector circuitry, consisting of the cables and preamplifiers (BW> 1.2 GHz) attached to the device itself, by the chip mount and on-chip components. These latter are expected to add a parasitic (simulated) capacitance of the order of 1 pF, which, in combination with the 5 kΩ resistance of the graphene channel, gives rise to a response time of a few ns, the largest limitation to our rise-time. In addition, the impinging THz QCL source is driven by a pulse generator, which has an intrinsic time jitter of 100 ps [68]. Moreover, in the present setup, the laser can undergo thermal fluctuations of the order of ±1 K during operation. This effect can smear out the pulses as a consequence of power fluctuations. The obtained 53 MHz BW has therefore to be considered as a lower limit. This, combined with the 160 pW/Hz$^{1/2}$ NEP, identifies our device as state-of-the-art amongst any other uncooled SLG THz detectors reported so far [11,29,36].

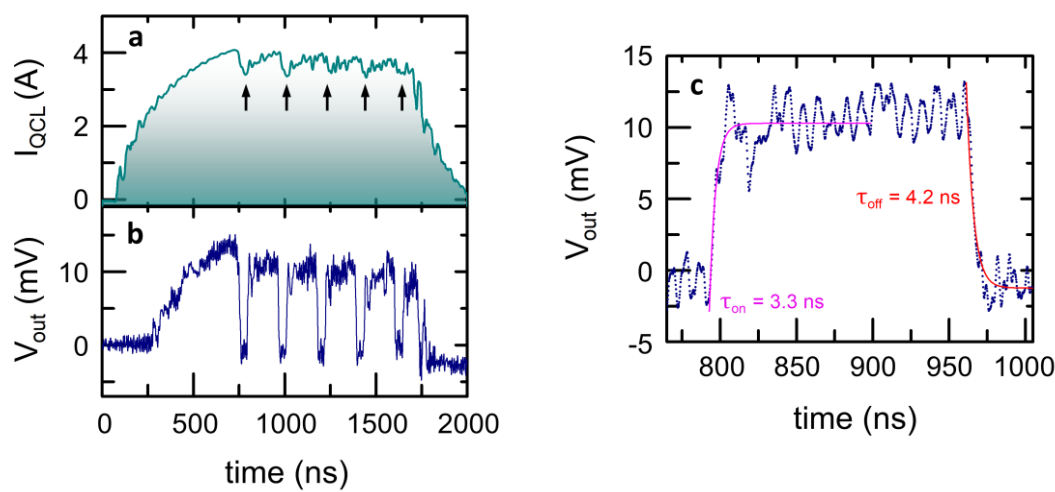

**Figure 4:** BW evaluation. (a) Driving current of the QCL when the laser is operated in pulse mode: pulse width 1.6 μs, repetition rate 33 Hz. At high bias ($V_{QCL} > 29$ V), the current presents fluctuations corresponding to the *on/off* switching of the laser. (b) Detector signal recorded at $V_G = 0$ V, with a 5.0 GS/s oscilloscope. The QCL turns off in an almost regular way (210 ns period). (c) Zooming between two consecutive fluctuations it is possible to retrieve the PD time constants by fitting the waveform with the standard charge-time, discharge-time expressions [36]. The estimated rise-time and fall-time are $\tau_{on} = 3.3$ ns and $\tau_{off} = 4.2$ ns, respectively, corresponding to BW ~ 53 MHz.

In conclusion, we demonstrate a record performance GFET PD with a 53 MHz modulation BW, operating at 3 THz. The device operation frequency is set by the coupling scheme, given here by a planar bow-tie antenna, and can be tailored across the whole THz range by engineering the antenna design. Changing the size of the antenna would indeed tune its resonance frequency and changing the type of antenna can narrow or broaden its frequency coverage. We took advantage of the peculiar power instabilities of the QCL source in specific transport regimes, to extract a response time $\tau$ ~ 3.3 ns. This, when combined with the achieved NEP, makes our PD the fastest, low noise, RT THz PD to date. We attribute the BW performance improvement to the high mobility of our hBN-SLG-hBN heterostructure. Further improvements are expected integrating our PDs with on-chip micro-strip lines, to reduce the overall circuit capacitance, therefore avoiding possible BW limitations induced by the external FET circuitry. Further refinements of the experimental system electronics (e.g. a larger BW, low noise amplifier) can help to assess the real intrinsic speed limit of our PDs, which is expected to be in the ~ 10 ps range [36].

**Supporting Information**

The Supporting Information includes: bow-tie antenna simulation, quantum cascade laser characterization, description of the photothermoelectric model. This material is available free of charge via the internet at http://pubs.acs.org.

**Acknowledgements**

We acknowledge funding from the ERC Consolidator Grant SPRINT (681379), the EU Graphene Flagship, ERC grant Hetero2D, EPSRC grants EP/L016087/1, EP/K01711X/1, EP/K017144/1. M.S.V. acknowledges partial support from the second half of the Balzan Prize 2016 in applied photonics delivered to Federico Capasso.